\documentclass[12pt,preprint]{aastex}

\begin{document}

\title{Directional Statistics for Polarization Observations of
Individual Pulses from Radio Pulsars}
\author{M. M. McKinnon}
\affil{National Radio Astronomy Observatory\altaffilmark{1}
\altaffiltext{1}{The National Radio Astronomy Observatory is a 
facility of the National Science Foundation operated under 
cooperative agreement by Associated Universities, Inc.} 
Socorro, NM, USA}

\begin{abstract}
Radio polarimetry is a three-dimensional statistical problem. The 
three-dimensional aspect of the problem arises from the Stokes 
parameters Q, U, and V, which completely describe the polarization 
of electromagnetic radiation and conceptually define the orientation 
of a polarization vector in the Poincar\'e sphere. The statistical
aspect of the problem arises from the random fluctuations in the
source-intrinsic polarization and the instrumental noise. A simple 
model for the polarization of pulsar radio emission has been used 
to derive the three-dimensional statistics of radio polarimetry. The 
model is based upon the proposition that the observed polarization 
is due to the incoherent superposition of two, highly polarized, 
orthogonal modes. The directional statistics derived from the model 
follow the Bingham-Mardia and Fisher family of distributions. The 
model assumptions are supported by the qualitative agreement between 
the statistics derived from it and those measured with polarization 
observations of the individual pulses from pulsars. The orthogonal 
modes are thought to be the natural modes of radio wave propagation 
in the pulsar magnetosphere. The intensities of the modes become 
statistically independent when generalized Faraday rotation (GFR) 
in the magnetosphere causes the difference in their phases to be 
large. A stochastic version of GFR occurs when fluctuations in the 
phase difference are also large, and may be responsible for the more 
complicated polarization patterns observed in pulsar radio emission.

\end{abstract}


\section{Introduction}

The four fundamental measurements made in astronomy are the intensity, 
flux density, or surface brightness of the electromagnetic radiation
emitted by a celestial object, the wavelength, or frequency, of the 
radiation, its location on the sky, and the polarization of the 
radiation. Measurements of the latter two, location and polarization,
follow the statistics of direction. The association of directional
statistics with the measurement of location is obvious, but the 
application of directional statistics to polarization measurements 
is not immediately apparent until one recalls that the Stokes parameters 
Q, U, and V describe the orientation of a polarization vector within 
the Poincar\'e sphere. The Stokes parameter V defines the circular
polarization of the radiation and establishes the \lq\lq z-coordinate"
of the polarization vector in the Poincar\'e sphere. The Stokes
parameters Q and U describe the radiation's linear polarization
and establish the vector's x- and y-coordinates, respectively. Here, 
polarization measurements are shown to follow directional statistics,
and these statistics are applied to polarization observations of radio 
pulsars.

Pulsars are rapidly rotating, highly magnetized neutron stars. Their 
rotation periods range between about 1ms and 10s, and the strength 
of the magnetic field at their surfaces ranges from $10^8$ G for the 
oldest pulsars to over $10^{12}$ G for the youngest. A beam of radio 
emission is emitted from each of the star's magnetic poles. A pulse 
of radio emission is observed as the star's rotation causes the beam 
to sweep across an observer's line of sight. Pulsar radio emission is 
generally thought to originate from charged particles streaming along 
open magnetic fields lines above the star's magnetic pole, but unlike 
other astrophysical radiative processes (e.g. synchrotron radiation, 
maser emission, and thermal radiation), it is poorly understood. 
Polarization observations of the individiual pulses from pulsars are 
made in an attempt to understand the radio emission mechanism and to 
study the propagation of radio waves in ultra-strong magnetic fields.

Polarization observations of individual pulses (Lyne et al. 1971;
Manchester et al. 1975; Backer \& Rankin 1980; Stinebring et al. 1984)
show that the radiation can be highly elliptically polarized and highly 
variable, if not stochastic. In many cases, the mean of the polarization 
position angle varies in an S-shaped pattern across the pulse. But 
histograms of position angle created from the single pulse observations 
show the angles follow the pattern in two parallel paths separated by 
about 90 degrees (Stinebring et al. 1984). Furthermore, histograms of 
fractional linear polarization show that the radiation is significantly 
depolarized at pulse locations where these orthogonally polarized (OPMs) 
modes occur. The OPMs are thought to be the natural modes of wave 
propagation in pulsar magnetospheres (Allen \& Melrose 1982; Barnard \& 
Arons 1986). The narrow bandwidths and short sampling intervals used in 
single pulse observations cause the instrumental noise in these 
observations to be large. The narrow bandwidths are used to overcome 
pulse smearing effects caused by the dispersion measure of, and multipath 
scattering in, the interstellar medium. The short sampling intervals, 
typically of order 100us, are needed to adequately resolve the short 
duration radio pulse. The combination of the stochastic nature of the 
intrinsic emission and the high instrumental noise suggests that a 
statistical approach is needed to analyze the single pulse data.

Most results from single pulse polarization observations have been 
reported as histograms of fractional linear polarization, fractional
circular polarization, and polarization position angle (Backer \& Rankin 
1980; Stinebring et al. 1984). While these display methods are extremely 
useful, they do not provide a complete picture of pulsar polarization 
because they force a separate interpretation of the circular and linear 
polarization, instead of a combined one as the observed elliptical 
polarization of the radiation would suggest. A complete, 
three-dimensional view of the polarization can be made by plotting the 
polarization measurements from a specific pulse location in the Poincar\'e 
sphere and projecting the result in two dimensions. The projections show 
how the orientation of the polarization vector fluctuates on the Poincar\'e 
sphere and reveal a wide variety of quasi-organized patterns. For example, 
in the cone emission at the edges of the pulse in PSR B0329+54 (Edwards \& 
Stappers 2004), the patterns consist of two clusters of data points, each 
in a separate hemisphere of the Poincar\'e sphere. In the precursor to the 
pulsar's central core component, the pattern is a single cluster of data 
points. Within the pulsar's core emission at the center of the pulse, one 
of the two clusters seen in the cone emission stretches into an ellipse or 
bar, while the other spreads into an intriguing partial annulus. The 
signatures of these patterns are not apparent in histograms of fractional 
polarization or position angle, emphazing the benefit of analyzing the 
Stokes parameters together. Any viable model of pulsar polarization must 
be able to replicate the observed patterns in addition to the histograms 
of fractional polarization. 

\section{Model of Pulsar Polarization}

The details of the statistical model for pulsar polarization are summarized 
in a series of papers by McKinnon and Stinebring (McKinnon \& Stinebring 
1998, 2000; McKinnon 2003, 2004, 2006, 2009). The main hypothesis of the 
model is the radiation's polarization is determined by the simultaneous 
interaction of two, highly polarized, orthogonal modes. By definition, 
the unit vectors representing the orthogonal modes are antiparallel in the 
Poincar\'e sphere and thus form a \lq\lq mode diagonal" in the sphere. The 
model accounts for the statistical nature of the observed polarization 
fluctuations by assuming the mode intensities are independent random 
variables. The assumption of statistical independence requires the 
difference in mode phases to be large (Melrose 1979) and greatly simplifies 
the model by allowing the mode intensities to be added (Chandrasekhar 1960). 
The model also accounts for the additive instrumental noise in each of the 
Stokes parameters. By assuming the mode intensities and instrumental noise 
are normal random variables, one can derive analytical expressions for the 
distributions of total intensity, polarization, and fractional polarization, 
as well as distributions for the orientation angles of the polarization 
vector. 

The main result from the model for the purposes of this paper is the 
derivation of the conditional density of the polarization vector's 
orientation angles. The conditional density is the joint probability 
density of the vector's colatitude, $\theta$, and longitude, $\phi$, 
at a fixed value of polarization amplitude, $r_o$. It captures the 
functional form of the more general joint density in a simple analytical 
expression. It is known as the Bingham-Mardia (Bingham \& Mardia 1978), 
or von Mises-Fisher, distribution.

\begin{equation}
f(\theta,\phi |r_o) = {\sin\theta\over{4\pi}}
{\exp[\pm\kappa^2(\cos\theta\pm\gamma)^2]\over{w(\kappa,\gamma)}}
\label{eqn:BM}
\end{equation}
The conditional density is parameterized by the constants $\kappa$ and 
$\gamma$ and is normalized by the constant $w$. The constant $\kappa$ 
can be regarded as a signal-to-noise ratio in polarization. The constant 
$\gamma$ satisfies the relation $|\gamma|\le 1$. By construction, the 
distribution is symmetric in longitude, which is uniformly distributed 
over $2\pi$. The vector's longitude and colatitude are statistically 
independent of one another. 

The plus signs in the argument of the exponential in 
Equation~\ref{eqn:BM} occur when the polarization fluctuations are 
predominantly parallel to the mode diagonal. They are caused by the 
randomly varying intensities of the OPMs. In this case, the functional 
form of the colatitude conditional density is generally bimodal. The 
polarization pattern formed by a projection of the conditional density
generally consists of a set of concentric circular contours in each 
hemisphere of the projection. The circular shape of the pattern arises 
from the symmetry in longitude.

The minus signs in the argument of the exponentail in 
Equation~\ref{eqn:BM} occur when the polarization fluctuations are 
predominantly perpendicular to the mode diagonal. The origin of these 
perpendicular fluctuations is not known, but is discussed in the 
following section. In this case, the conditional density is always 
unimodal because it is normal in $\cos\theta$. The polarization pattern
formed by the projection of this conditional density is generally a 
complete annulus in only one of the two projection hemispheres (McKinnon 
2009).

The general applicability of Equation~\ref{eqn:BM} can be illustrated
with a few special cases. When $\kappa=0$, the polarization fluctuations 
are dominated by instrumental noise, and the conditional density becomes 
isotropic, as one would expect for pure noise. When $\kappa\gg 1$, the 
fluctuations are very small in comparison to the polarized signal, and 
the conditional density becomes a Fisher distribution (Fisher et al. 1987). 
When $\gamma=0$ and the fluctuations are predominantly along the mode 
diagonal, as caused by OPMs, the mean intensities of the modes are equal, 
the modes occur with equal frequency, and the conditional density 
becomes the Watson bipolar distribution (McKinnon 2006; Fisher et al. 1987).

The joint probability density of the vector's colatitude and longitude 
has been shown to be a reasonable representation of the distribution 
of angles that are actually observed (McKinnon 2006). The conditional
density has been shown to produce projections of the Poincar\'e sphere
that are qualitatively consistent with the polarization patterns observed 
in pulsar radio emission (McKinnon 2009).

\section{Generalized Faraday Rotation}

Two aspects of the model and its application to the observations 
require additional explanation. These are (1) an explanation for the 
mechanism that causes the difference in mode phases to be large, thereby 
providing additional justification for the assumption of independent 
mode intensities and (2) a physical explanation for the mechanism that 
creates the fluctuations perpendicular to the mode diagonal, which were
incorporated in the model to account for annular polarization patterns. 
The explanation for both may reside with generalized Faraday rotation 
(GFR; Edwards \& Stappers 2004; McKinnon 2009).

In general terms, Faraday rotation is the physical process that alters 
the difference between the phases of the modes as they propagate through 
a plasma (Melrose 1979). The modes are incoherent when the difference in 
their phases {\it at a given wavelength} is large ($\Delta\chi\gg 1$) and 
are coherent (coupled) as long as the phase difference is small 
($\Delta\chi < 1$). The modes retain their individual polarization 
identity in an observation when they are incoherent, but effectively 
lose their individual identity when they are coherent. Faraday rotation 
can become stochastic when the fluctuations in phase difference are large 
($\sigma_\chi \gg 1$; Melrose \& Macquart 1998).

GFR alters the component of the radiation's polarization vector that is
perpendicular to the polarization vectors of the plasma's wave propagation
modes. For any plasma, the unit vectors representing the polarization 
states of the two modes are anti-parallel on a diagonal through the 
Poincar\'e sphere. For the cold, weakly-magnetized plasma that is the 
interstellar medium (ISM), the propagation modes are circularly polarized, 
and the mode diagonal defined by their polarization vectors connects the 
poles of the Poincar\'e sphere. Faraday rotation in the ISM causes the 
orientation of the radiation's polarization to vary in a plane 
perpendicular to the mode diagonal, either on the Poincar\'e sphere's 
equator or on a small circle parallel to it, depending upon the 
polarization state of the plasma-incident radiation. For the relativistic 
plasma in the strong magnetic field of a pulsar's magnetosphere, the modes 
are thought to be linearly polarized (Allen \& Melrose 1982; Barnard \& 
Arons 1986; Melrose 1979) so that the mode diagonal lies in the equatorial 
plane of the Poincar\'e sphere. Similar to Faraday rotation in the ISM, 
GFR in a pulsar's magnetosphere causes the polarization vector to rotate 
on a small circle in the Poincar\'e sphere that is perpendicular to and 
centered on the mode diagonal (e.g. see Fig. 3 of Kennett \& Melrose 1998). 
Random fluctuations in $\Delta\chi$ (i.e. stochastic GFR) would appear as 
a partial annulus around the mode diagonal, as is observed in the core 
component of PSR B0329+54 (Edwards \& Stappers 2004).

\begin{figure}
\plotone{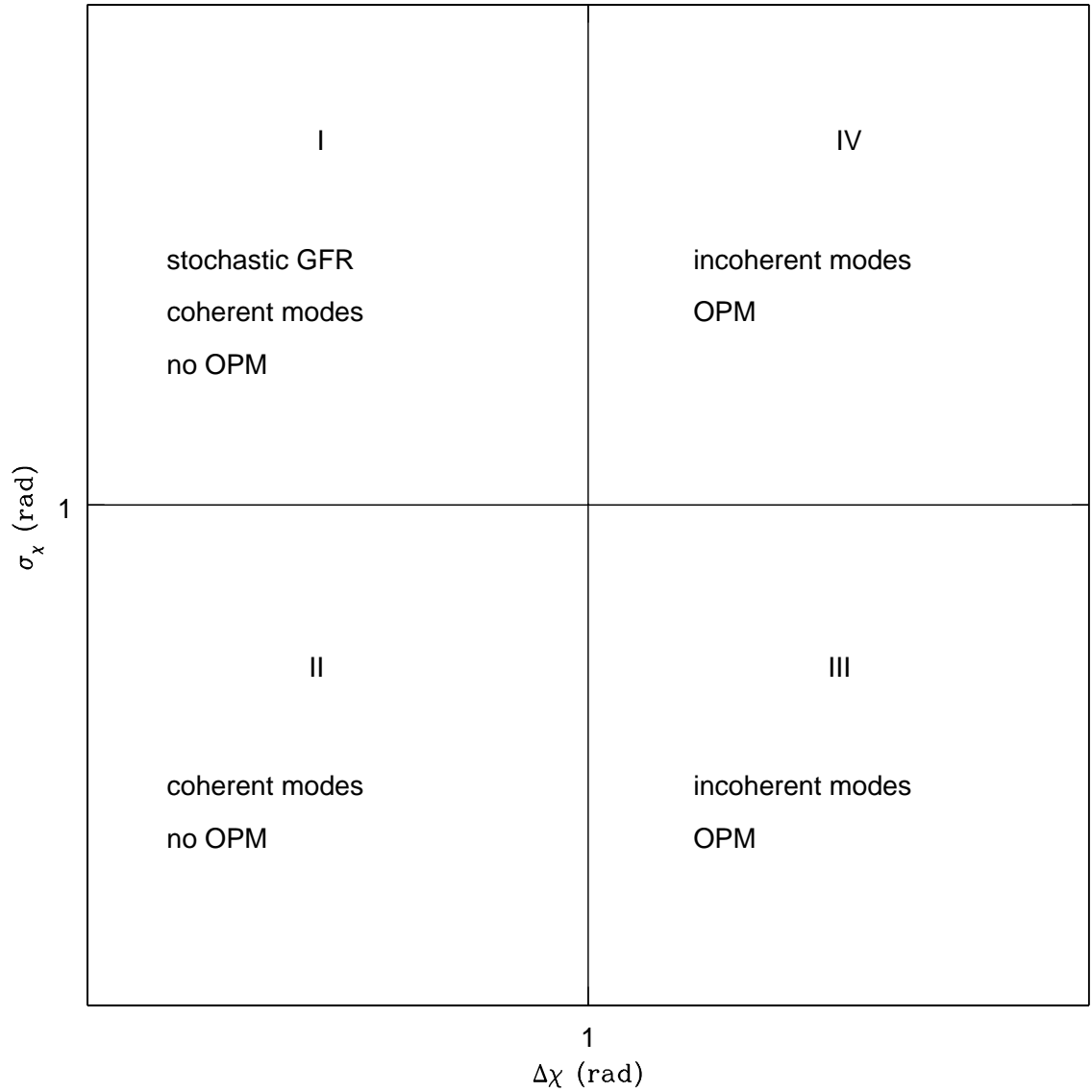}
\caption{Requirements on mode phase difference, $\Delta\chi$, and its 
fluctuations, $\sigma_\chi$, for OPM and stochastic GFR to occur.}
\label{fig:GFR}
\end{figure}

Figure~\ref{fig:GFR} is a plot of $\sigma_\chi$ versus $\Delta\chi$ and
summarizes the discussion above. The plot is divided into four regions,
I through IV, that define the conditions under which OPM and stochastic
GFR can occur. OPMs can occur only in regions III and IV, to the 
right of $\Delta\chi=1$, where the modes are incoherent. The modes
are coherent when $\Delta\chi<1$; therefore, OPMs will not be
observed when conditions in the pulsar magnetosphere (or the ISM) are 
consistent with those in regions I and II. The Faraday rotation that is 
typically observed in the ISM or in the lobes of extragalactic radio jets 
occurs in region II where the modes are coherent, but the fluctuations 
in $\Delta\chi$ are small. Stochastic GFR can occur only under the 
conditions specific to region I, where the modes are coherent but the 
fluctuations in $\Delta\chi$ are large. Returning now to the observations,
the bimodal polarization pattern observed in the cone emission of PSR 
B0329+54 arises from OPMs. The mean and standard deviation of 
$\Delta\chi$ at this location of the pulse would reside in region III 
or IV of Figure~\ref{fig:GFR}. The properties of $\Delta\chi$ in the 
pulsar's core precursor also likely reside in region III or IV of the 
figure, even though the polarization pattern at this pulse location 
consists of a single cluster of data points. OPMs clearly occur everywhere 
else within the pulse. OPMs may also occur in the precursor, but one of 
the modes may be so strong that the other mode is never detected. However, 
one cannot rule out the possibility that the properties of $\Delta\chi$ 
in the precursor reside within region II of the figure. The polarization 
pattern in the core component of PSR B0329+54 is much more complicated 
because both modes are present, but one of them reveals itself as a 
partial annulus. The statistical model decribed here cannot completely
explain this behavior. The pattern may arise from a condition that falls 
on the regional boundaries of Figure~\ref{fig:GFR}, where the modes are 
occasionally coherent with large fluctations in $\Delta\chi$ (i.e. in 
region I of the figure), thus explaining the partial annulus, but are 
otherwise incoherent (i.e. in region III or IV) to account for the 
bimodal aspect of the polarization pattern.

\section{Conclusions}

A statistical model has been developed for the polarization of pulsar 
radio emission. The model can explain a wide variety of polarization
patterns observed in the radio emission. The observations are thus
consistent with the model's hypothesis that the polarization of the 
radiation is determined by the simultaneous interaction of two, highly 
polarized, orthogonal modes. The analysis of the polarization data 
shows that polarization signatures of physical processes can become 
apparent when the Stokes parameters are analyzed together, instead of 
separately. An interpretation of the model's assumptions and its 
application to the observations suggest that generalized Faraday 
rotation may be operative in pulsar magnetospheres. The model shows, 
in a rigorous way, that polarization measurements follow the statistics 
of direction.


\begin{references}
\reference{} Allen, M. C. \& Melrose, D. B. 1982, Proc. Astron. Soc.
             Aust., 4, 365
\reference{} Backer, D. C. \& Rankin, J. M. 1980, \apjs, 42, 143
\reference{} Barnard, J. J. \& Arons, J. 1986, \apj, 302, 138
\reference{} Bingham, C. \& Mardia, K. V. 1978, Biometrika, 65, 379
\reference{} Chandrasekhar, S. 1960, Radiative Transfer, 
             (New York: Dover)
\reference{} Edwards, R. T. \& Stappers, B. W. 2004, A\&A, 421, 681
\reference{} Fisher, N. I, Lewis, T., \& Embleton, B. J. J. 1987,
             Statistical Analysis of Spherical Data, (Cambridge: 
             Cambridge)
\reference{} Kennett, M. \& Melrose, D. 1998, Proc. Astron. Soc. Aust., 
             15, 211
\reference{} Lyne, A. G., Smith, F. G., \& Graham, D. A. 1971, \mnras,
             153, 337
\reference{} Manchester, R. N., Taylor, J. H., \& Huguenin, G. R. 1975,
             \apj, 196, 83
\reference{} Melrose, D. B. 1979, Aust. J. Phys., 32, 61
\reference{} Melrose, D. B. \& Macquart, J.-P. 1998, \apj, 505, 921
\reference{} McKinnon, M. M. 2003, \apjs, 148, 519
\reference{} McKinnon, M. M. 2004, \apj, 606, 1154
\reference{} McKinnon, M. M. 2006, \apj, 645, 551
\reference{} McKinnon, M. M. 2009, \apj, 692, 459
\reference{} McKinnon, M. M. \& Stinebring, D. R. 1998, \apj, 502, 883
\reference{} McKinnon, M. M. \& Stinebring, D. R. 2000, \apj, 529, 435
\reference{} Stinebring, D. R. et al., 1984, \apjs 55, 247

\end{references}
\end{document}